\documentclass[%
 reprint,
 amsmath,amssymb,
 aps,
]{revtex4-1}
\usepackage{graphicx}% Include figure files
\usepackage{dcolumn}% Align table columns on decimal point
\usepackage{bm}% bold math
\begin{document}

\title{Band structure of Ge$_{1-x}$Sn$_{x}$ alloy: a full-zone 30-band $k$$\cdot$$p$ model}%

\author{Zhigang Song$^{1}$, Weijun Fan$^{1}$, Chuanseng Tan$^{1}$, Qijie Wang$^{1}$, Donguk Nam$^{1}$, Daohua Zhang$^{1}$ and Greg Sun$^{2}$}
\affiliation{$^{1}$School of Electrical and Electronic Engineering, Nanyang Technological University, 50 Nanyang Avenue, Singapore 639798, Singapore}
\affiliation{$^{2}$Department of Engineering,
University of Massachusetts Boston, Massachusetts 02125, U.S.A}

\date{\today}

\begin{abstract}
A full-zone 30-band $k$$\cdot$$p$ model is developed as an efficient and reliable tool to
compute electronic band structure in Ge$_{1-x}$Sn$_{x}$ alloy. The model was first used to
reproduce the electronic band structures in Ge and $\alpha$-Sn obtained with empirical
tight binding and \textit{ab initio} methods. Input parameters for the 30-band $k$$\cdot$$p$ model are
carefully calibrated against prior empirical predications and experimental data.
Important material properties such as effective mass for electrons and holes,
Luttinger parameters, and density of states are obtained for Ge$_{1-x}$Sn$_{x}$ alloy with the
composition range $0<x<0.3$. The 30-band $k$$\cdot$$p$ model that requires far less computing
resources is a necessary capability for optimization of sophisticated devices made
from Ge$_{1-x}$Sn$_{x}$ alloy with a large parameter space to explore.

\end{abstract}

\maketitle

\section{Introduction}

 The field of Si photonics has seen impressive growth since the early vision in 1990s \cite{soref1993silicon}. The huge infrastructure of the global Si electronics industry is expected to benefit the fabrication of highly sophisticated Si photonic devices at costs that are lower than those currently required for compound semiconductors. Following the discovery of high speed SiGe electronics \cite{meyerson1994high}, the landscape of Si-based optical platform quickly expanded to include Ge and SiGe alloys. Among all the necessary components that make up the complete set of the Si photonics, efficient light sources such as LEDs and lasers are the most challenging because Si, Ge, and their alloys are all indirect bandgap materials. In order to improve the emission efficiency, effort has been directed towards developing direct bandgap material by incorporating yet another group-IV element, Sn, into the mix \cite{soref1996silicon}. Early theoretical studies suggested that the group-IV alloy SiGeSn may possess a direct bandgap since Sn has negative bandgap at $\Gamma$-point \cite{soref1993direct}. But the progress in developing actual light sources has been painfully slow due to the grand technical challenges in growing high quality SiGeSn materials with sufficient Sn to turn the materials into direct bandgap. Si, with its band gap at $\Gamma$ valley sitting 2.28 eV above its indirect band gap at the X-valley \cite{madelung2012semiconductors}, naturally requires high Sn composition to pull its $\Gamma$-valley bandgap down to form direct-bandgap SiSn alloy. The difference for Ge, on the other hand, is smaller - only 0.14 eV between its indirect L-valley and direct $\Gamma$ valley \cite{shur1996handbook}. As a result, relatively small amount of Sn composition, estimated to be around $6-8\%$\cite{jenkins1987electronic}, is sufficient for GeSn alloy to turn direct. Mid infrared (IR) electrical-injection lasers based on direct-bandgap GeSn as active region with latticed matched SiGeSn barrier have been predicted \cite{sun2010design1,sun2010design2}. To date optically pumped mid-IR GeSn lasers have been demonstrated \cite{stange2016optically,al2016optically,dou2018optically}. There is reason to be optimistic that the grand prize this material system has to offer, i.e., electrically pumped GeSn laser diodes, is finally within reach and the community will soon be facing the task of device performance optimization. This task is complicated by the fact that many material parameters remain unknown for a wide range of alloy compositions that need to be extracted from electronic band structures across the entire Brillion zone (BZ).

The electronic band structure of bulk Ge$_{1-x}$Sn${_{x}}$ alloy can be calculated using empirical pseudopotential method (EPM) \cite{yahyaoui2014wave,lu2012electronic, moontragoon2012direct}, empirical tight binding (ETB) \cite{attiaoui2014indirect} and \textit{ab initio} \cite{zelazna2015electronic, polak2017electronic} approaches. These methods are known to be accurate in predicting electronic band structures that agree well with experiments \cite{polak2017electronic}, but the demand on computing time and resource makes them impractical to be employed in optimization of sophisticated photonic devices that involve heterostructures and/or nanostructure such as superlattices, quantum wells and quantum dots, especially when a large parameter space needs to be explored. In comparison, the 8-band $k$$\cdot$$p$ model is far more efficient in calculating the band structure near the $\Gamma$-point of the BZ, which works well for direct bandgap III-V semiconductors since essentially all electronic and optical processes occur close to the proximity of the $\Gamma$-point. However, for GeSn alloys where the $\Gamma$ and L valleys are not separated far apart in energy, electrons are subject to inter-valley scattering such that both valleys are populated. It is therefore necessary to obtain full BZ band structure with a model that is computationally efficient. The 30-band $k$$\cdot$$p$ model developed as an expansion of the simpler 8-band approach is capable of calculating the full BZ band efficiently and has been used successfully in reproducing full-zone band structures for Si, Ge, SiGe alloy \cite{rideau2006strained}, $\alpha$-Sn \cite{PhysRevB.2.352} and other III-V structures \cite{boyer201130}.

In this article, we present a 30-band $k$$\cdot$$p$ model that can be used to map out full-zone electronic energy band structures for Ge$_{1-x}$Sn$_{x}$ alloys where Sn composition varies from 0 to 0.3. This composition range covers most GeSn alloys currently under development for device
applications. The model is anchored at reproducing the previously known band structures of Ge and $\alpha$-Sn that are obtained using either ETB or \textit{ab initio} method. This step establishes a set of reliable input parameters for the 30-band $k$$\cdot$$p$ model at the two extreme compositions: $x=0, 1$. The parameters needed for any Ge$_{1-x}$Sn$_{x}$ alloy in $0<x<0.3$ are then obtained in the spirit of virtual crystal approximation (VCA) with either linear or quadratic interpolations between the two extreme points at the composition $x$ that is adjusted with available experimental results. The electronic band structures calculated with the 30-band $k$$\cdot$$p$ model allow us to extract many electronic and optical properties that are important to electronic and photonic devices of all kinds made from Ge$_{1-x}$Sn$_{x}$. In this work we present the results of effective masses along different crystalline directions, Luttinger parameters, and density of states (DOS) at $\Gamma$ and L valleys for relaxed Ge$_{1-x}$Sn$_{x}$ alloys spanning across the alloy composition range $0.0<x<0.3$ at room temperature. In addition, this model can be used to pinpoint the alloy composition $x$ for Ge$_{1-x}$Sn$_{x}$ to transition from indirect to direct bandgap and the prediction $x=7.25\%$ is in good agreement with experimental measurement \cite{tseng2013mid}. It should be pointed out that for the purpose of demonstration we choose to study relaxed GeSn at room temperature, but this model can easily be extended to include the effect of strain as well as at other temperatures. This study illustrates the efficiency and effectiveness of the 30-band $k$$\cdot$$p$ model in calculating the electronic band structures of the Ge$_{1-x}$Sn$_{x}$ alloy, allowing for extraction of practically all electronic and optical properties associated with its energy band for device simulation and optimization.

\section{Theoretical model}
\begin{figure}
\includegraphics[width=3.0in]{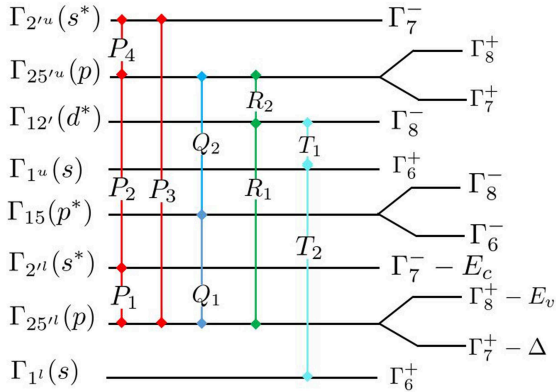}
\caption{Schematic of the zone centre energy levels with symmetry classification in the double group representation.}
\label{fig1}
\end{figure}

%Using a zinc-blende $\Gamma$-centered Bloch function basis
%function basis $u_{lk}(\mathbf{r})=\sum_{n}C_{n}^{l}u_{n0}(\mathbf{r})$, the $k$$\cdot$$p$ equation of ideal crystal is:
%\begin{small}
%\begin{equation}
%   \sum_{n}\left\{ \left( \frac{\hbar ^{2}k^{2}}{2m}+E_{n}^{0}-E_{lk}\right) \delta
%_{n,n^{\prime }}+\frac{\hbar \mathbf{k}}{m}\cdot \left\langle u_{n^{\prime
%}0}\left\vert \mathbf{p}\right\vert u_{n0}\right\rangle \right\} C_{n}^{l}=0
%\end{equation}
%\end{small}
%where $E_{n}^{0}$ are the eigenvalues at $\Gamma$.
The 30-band $k$$\cdot$$p$ model Hamiltonian matrix can be written as \cite{rideau2006strained}:
\begin{widetext}
\begin{equation}\label{eq1}
    H_{kp}^{30}=\left(
\begin{array}{cccccccc}
H_{\Gamma _{2^{\prime u}}}^{2\times 2} & P_{4}H_{k}^{2\times 6} & 0 & 0 & 0
& 0 & 0 & P_{3}H_{k}^{2\times 6} \\
& H_{\Gamma _{25^{\prime u}}}^{6\times 6} & R_{2}H_{k}^{6\times 4} & 0 & 0 &
Q_{2}H_{k}^{6\times 6} & P_{2}H_{k}^{6\times 2} & H_{\Gamma _{25^{\prime
u}}\Gamma _{25^{\prime l}}}^{SO} \\
&  & H_{\Gamma _{12^{\prime }}}^{4\times 4} & 0 & 0 & 0 & 0 &
R_{1}H_{k}^{4\times 6} \\
&  &  & H_{\Gamma _{1^{u}}}^{2\times 2} & 0 & T_{1}H_{k}^{2\times 6} & 0 & 0
\\
&  &  &  & H_{\Gamma _{1^{l}}}^{2\times 2} & T_{2}H_{k}^{2\times 6} & 0 & 0
\\
&  &  &  &  & H_{\Gamma _{15}}^{6\times 6} & 0 & Q_{1}H_{k}^{6\times 6} \\
&  &  &  &  &  & H_{\Gamma _{2^{\prime l}}}^{2\times 2} &
P_{1}H_{k}^{2\times 6} \\
&  &  &  &  &  &  & H_{\Gamma _{25^{\prime l}}}^{6\times 6}%
\end{array}%
\right)
\end{equation}
 \end{widetext}

whose diagonal blocks of different orders
\begin{equation}
\begin{split}
H_{\Gamma }^{6\times 6} &=\text{diag}(E_{\Gamma }+\frac{\hbar ^{2}k^{2}}{2m})+H_{\Gamma }^{SO} \\
H_{\Gamma }^{4\times 4} &=\text{diag}(E_{\Gamma }+\frac{\hbar ^{2}k^{2}}{2m}) \\
H_{\Gamma }^{2\times 2} &=\text{diag}(E_{\Gamma }+\frac{\hbar ^{2}k^{2}}{2m}
)
\end{split}
\end{equation}
where the wavevector $k^{2}=k_{x}^{2}+k_{y}^{2}+k_{z}^{2}$, and diag$(...)$ stands for diagonal matrix. $E_{\Gamma}$ is the eigen energy level for the considered $\Gamma$ states. $H^{SO}_{\Gamma}$ is the spin-orbital coupling (SOC) matrix:
\begin{equation}
H_{\Gamma }^{SO}=\frac{\Delta _{\Gamma }}{3}\left(
\begin{array}{cccccc}
-1 & -i & 0 & 0 & 0 & 1 \\
i & -1 & 0 & 0 & 0 & -i \\
0 & 0 & -1 & -1 & i & 0 \\
0 & 0 & -1 & -1 & i & 0 \\
0 & 0 & -i & -i & -1 & 0 \\
1 & i & 0 & 0 & 0 & -1%
\end{array}%
\right)
\end{equation}
where ${\Delta _{\Gamma }}$ is the SOC strength. The nonzero off-diagonal blocks in Eq. (\ref{eq1}) are given as
\begin{subequations}

\begin{equation}
H_{k}^{6\times 6}=\left(
\begin{array}{cccccc}
0 & k_{z} & k_{y} & 0 & 0 & 0 \\
k_{z} & 0 & k_{x} & 0 & 0 & 0 \\
k_{y} & k_{x} & 0 & 0 & 0 & 0 \\
0 & 0 & 0 & 0 & k_{z} & k_{y} \\
0 & 0 & 0 & k_{z} & 0 & k_{x} \\
0 & 0 & 0 & k_{y} & k_{x} & 0%
\end{array}%
\right)
\end{equation}

\begin{equation}
H_{k}^{4\times 6}=\left(
\begin{array}{cccccc}
0 & \sqrt{3}k_{y} & -\sqrt{3}k_{z} & 0 & 0 & 0 \\
2k_{x} & -k_{y} & -k_{z} & 0 & 0 & 0 \\
0 & 0 & 0 & 0 & \sqrt{3}k_{y} & -\sqrt{3}k_{z} \\
0 & 0 & 0 & 2k_{x} & -k_{y} & -k_{z}%
\end{array}%
\right)
\end{equation}

\begin{equation}
H_{k}^{2\times 6}=\left(
\begin{array}{cccccc}
k_{x} & k_{y} & k_{z} & 0 & 0 & 0 \\
0 & 0 & 0 & k_{x} & k_{y} & k_{z}%
\end{array}%
\right)
\end{equation}

\end{subequations}
The 15 energy-level parameters of various $\Gamma$ states from $O^{h}$ and the SOC strength introduced by Condona and Pollock \cite{PhysRev.142.530} can be obtained
from theoretical methods such as ETB, MBJLDA and HSE06.
The other ten parameters, $P_{1}, P_{2}, P_{3}, P_{4}, Q_{1}, Q_{2}, R_{1}, R_{2}, T_{1}, T_{2}$, used in the present 30-band $k$$\cdot$$p$ model are the matrix elements of the linear
 momentum operator $\mathbf{p}$, such as $P_{1}= \frac{\hbar }{m}\left\langle \Gamma _{25^{\prime l}}\left\vert
\mathbf{p}\right\vert \Gamma _{2^{\prime l}}\right\rangle$, representing the coupling strength between different bands as shown in Fig. \ref{fig1}, which can be optimized by nonlinear methods.

Based on above 30-band $k$$\cdot$$p$ model, effective mass can be calculated by the finite difference methods, where the second and mixed derivatives are evaluated using five-point stencil. In addition, to accelerate the speed of computation and save computation resources in the DOS calculation, sampling BZ methods \cite{PhysRevB.13.5188} is used. The k-points and corresponding weights are obtained through the Vienna $ab$ $initio$ simulation package (VASP )\cite{PhysRevB.54.11169}.

\section{Results and Discussion}
\begin{figure}
\includegraphics[width=3.4in,height=3.4in]{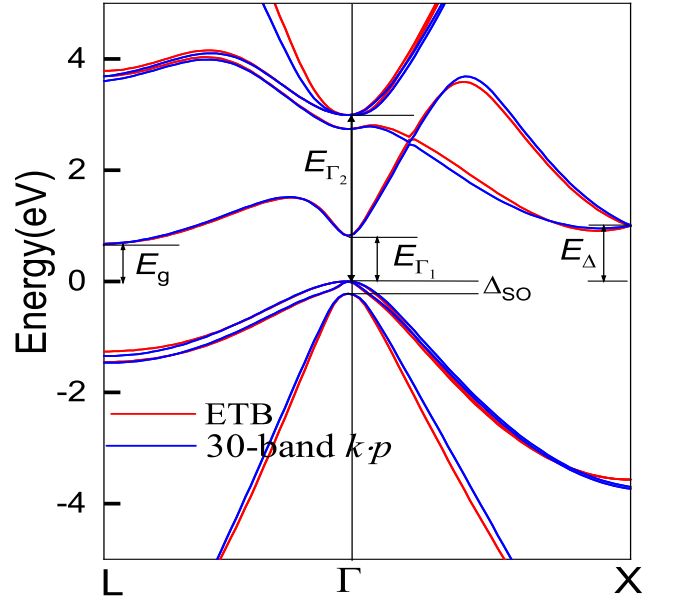}
\caption{Ge electronic band structure obtained from the ETB method(red) and fitted by the 30-band $k$$\cdot$$p$(blue) at room temperature.}
\label{fig2}
\end{figure}

The starting point of our calculation is the electron structure of single crystal Ge and $\alpha$-Sn.
The previous application of the 30-band $k$$\cdot$$p$ model to obtain the Ge band structure \cite{rideau2006strained} was based on parameters at liquid helium temperature. As we know the band structure of Ge has a strong dependence on temperature, for instance, the Ge band gaps at L-valley and $\Gamma$-valley are 0.66 eV and 0.80 eV, respectively, at room temperature which are significantly lower than those at 0.747 eV and 0.89 eV  at that of liquid helium \cite{madelung2012semiconductors,shur1996handbook}.
Since essentially all GeSn-based devices are developed for room temperature operation, we recalculated the Ge band structure at room temperature using the ETB method and then extracted parameters from it to feed into the 30-band $k$$\cdot$$p$ model in order to reproduce the band structure again. For convenience, we assume that the the valence band maximum(VBM) is at potential zero and that all other values are referenced to it in all calculations. Results from the two different methods are shown in Fig. \ref{fig2} where the two band structures are nicely matched globally across the full BZ. Table \ref{tab:table1} lists eigenvalues, energy gaps, effective masses and Luttinger parameters at high-symmetry points in BZ that are obtained from experiments, ETB and the 30-band $k$$\cdot$$p$ methods at 300K for Ge. The good overall agreement between them clearly validates the effectiveness of the 30-band $k$$\cdot$$p$ model.

\begin{figure}
\includegraphics[width=3.4in]{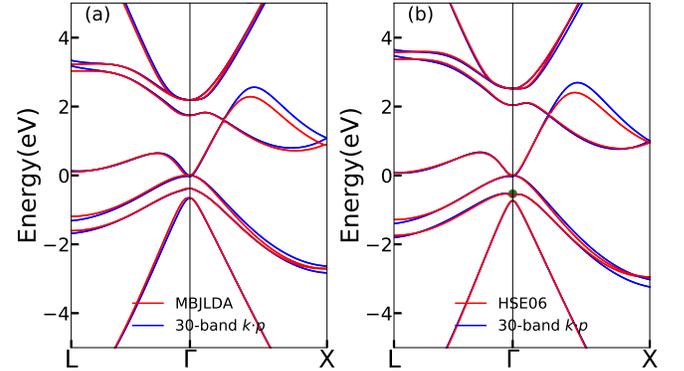}
\caption{$\alpha$-Sn electronic band structure obtained using (a) MBJLDA and (b) HSE06 hybrid functional calculation in red. Results by the 30-band $k$$\cdot$$p$ model calculation with input parameters extracted from respective band structures are shown in blue.}
\label{fig3}
\end{figure}

For $\alpha$-Sn it has been reported that \textit{ab initio} calculation could result in underestimating the band gap because of its negative band gap \cite{PhysRevB.2.352}, we therefore chose to employ two
amended methods involving hybrid functional to improve the \textit{ab initio} calculation. Two such methods are used to obtain the electronic band structure of $\alpha$-Sn. One is based on modified Becke-Johnson local density approximation (MBJLDA) \cite{becke2006simple} and result is shown (red curve) in Fig. \ref{fig3}(a). The other is proposed by Heyd, Scuseria, and Ernzerhof (HSE06) \cite{1.1564060} and its result is shown (red curve) in Fig. \ref{fig3}(b). From the two band structures, we have extracted two sets of parameters to feed into the 30-band $k$$\cdot$$p$ model. In both cases, they were able to reproduce the electronic band structure shown (blue curve) in Fig. \ref{fig3} that matches the two \textit{ab initio} hybrid functional results reasonably well, except near the X-valley where a substantial discrepancy can be observed. While this discrepancy remains to be resolved in future investigation, it does not affect the material properties that we aim to extract because $\Gamma$-valley and L-valley in conduction band are the major concerns for GeSn alloys as they sit significantly lower than the X-valley in energy.
It is not difficult to see from Fig. \ref{fig3} that although the
two different hybrid functional methods have produced largely matching electronic band structure
over the entire BZ, there are some inconsistencies. In the case of HSE06 method, for instance, the effective mass calculated from the $\Gamma_7$ conduction band is
positive at $\Gamma$-point as shown in Fig. \ref{fig3}(b) (green circle), but its experimental value as measured
turns out to be negative \cite{madelung2012semiconductors,shur1996handbook} which agrees with what was obtained from the MBJLDA method. The Similar discrepancy has aslo been found in HgTe and ScPtBi topological material with the HSE06 approach \cite{chadov2010tunable}.
%(In addition, another essential term is the inversion between $\Gamma_{25'^{l}}$ and $\Gamma_{2'}^{l}$, which has been confirmed in theory and experiments \cite{madelung2012semiconductors}. It also will be reflected in parameters in Table \ref{tab:table3}.)
The general agreement between the two methods and the possibility of local inconsistency with the HSE06 method lead us with confidence to follow the MBJLDA method for the subsequent
calculation of the electronic band structures of Ge$_{1-x}$Sn$_{x}$ alloys and extraction of their properties.

%grey tin band structure\cite{Sn_band_expt,Short-wave}

\begin{table}
 \caption{\label{tab:table1}Effective masses, Luttinger parameters and energy gaps obtained from experiments, ETB and $k$$\cdot$$p$ methods at 300K for Ge. All energies are in eV and the effective mass unit is $m_{0}$ that is the mass of free electron. Corresponding experiments values are from Refs. \cite{madelung2012semiconductors} and \cite{shur1996handbook}.}

\begin{ruledtabular}
\begin{tabular}{cccc}
Ge &Expt. &ETB  &$kp$ \\ \hline
$m_{e}^{\Gamma}$&0.038 \footnotemark[1]& 0.038& 0.045 \\
$m_{l}^{L}$& 1.59\footnotemark[1]& 1.588& 1.544 \\
$m_{t}^{L}$& 0.082\footnotemark[1]& 0.081& 0.085 \\
$m_{hh}^{001}$& 0.284\footnotemark[1]& 0.173& 0.194 \\
$m_{hh}^{110}$& 0.352\footnotemark[1]& 0.368& 0.399 \\
$m_{hh}^{111}$& 0.376\footnotemark[1]& 0.531& 0.558 \\
$m_{lh}^{001}$& 0.044\footnotemark[1]& 0.049& 0.058 \\
$m_{lh}^{110}$& 0.043\footnotemark[1]& 0.042& 0.050 \\
$m_{lh}^{111}$& 0.043\footnotemark[1]& 0.041& 0.048 \\
$\gamma_{1}$& 13.35\footnotemark[1]& 13.130& 11.298 \\
$\gamma_{2}$& 4.25\footnotemark[1]& 5.063& 3.153 \\
$\gamma_{3}$& 5.69\footnotemark[1]& 3.466& 4.751 \\
$E_{g}$& 0.66 \footnotemark[2]& 0.678& 0.670 \\
$E_{X}$& 1.2\footnotemark[2]& 0.998& 1.000 \\
$E_{\Gamma_{1}}$& 0.80\footnotemark[2]& 0.814& 0.814 \\
$E_{\Gamma_{2}}$& 3.22\footnotemark[2]& 2.990& 2.990 \\
$E_{\Delta}$& 0.85\footnotemark[2]& 0.905& 0.952 \\
$\Delta_{SO}$& 0.29\footnotemark[2]& 0.225& 0.225 \\

\end{tabular}
\end{ruledtabular}

\footnotetext[1]{From Ref.~\onlinecite{madelung2012semiconductors}.}
\footnotetext[2]{From Ref.~\onlinecite{shur1996handbook}.}
\end{table}

\begin{figure}[htb]
\includegraphics[width=3.4in]{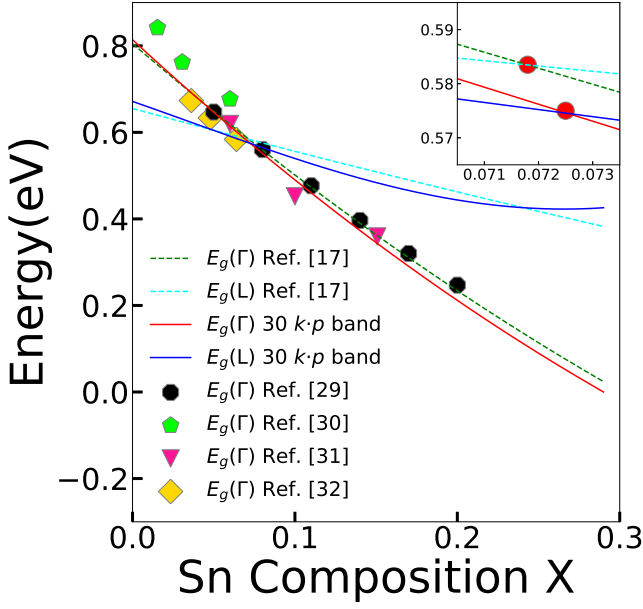}
\caption{The bandgaps at L (solid blue) and $\Gamma$-valley (solid red) vs Sn composition. The dashed curves are bandgaps at L (dashed light blue) and  $\Gamma$-valley (dashed green) according to the empirical quadratic interpolation \cite{zelazna2015electronic}. The scatter points are experimental data \cite{PhysRevB.73.125207,dybala2016electromodulation,PhysRevB.77.073202,lin2012investigation}. The zoom inset
shows the indirect-direct crossover at the Sn composition of $7.25\%$ for our 30-band $k$$\cdot$$p$ model compared with that of $7.18\%$ from Ref. \cite{zelazna2015electronic}.
 }
\label{fig4}
\end{figure}

\begin{figure}[htbp]
\includegraphics[width=3.4in,height=3.0in]{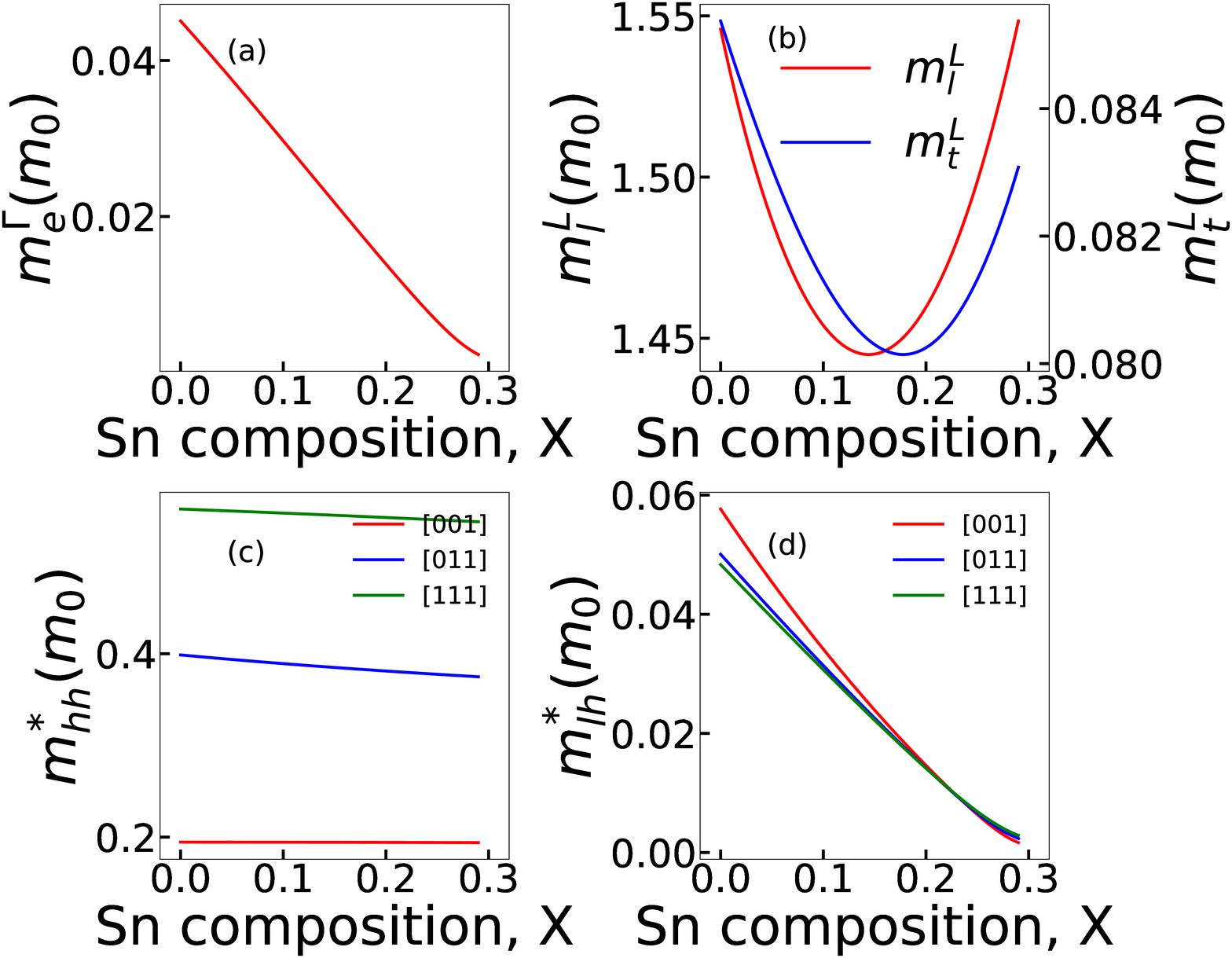}
\caption{(a) Effective mass of electron at $\Gamma$-valley. (b) Longitudinal and transverse electron effective masses at L valley.
(c) Heavy-hole effective masses along [001], [011], and [111] directions.
(d) Light-hole effective masses along  [001], [011], and [111] directions.}
\label{fig5}
\end{figure}

\begin{table}
\caption{\label{tab:table2}Matrix elements of the linear momentum $\mathbf{p}$(a.u.) used in the present 30-band $k$$\cdot$$p$ model. All are in atomic unit(a.u.).}
\begin{ruledtabular}
\begin{tabular}{cc}
Matrix elements(a.u.) &Ge$_{1-x}$Sn$_{x}$ \\ \hline
$P_{1}= \frac{\hbar }{m}\left\langle \Gamma _{25^{\prime l}}\left\vert
\mathbf{p}\right\vert \Gamma _{2^{\prime l}}\right\rangle$ & 1.1701+0.4655$x$ - 2.2767$x^{2}$\\
$P_{2}=\frac{\hbar }{m}\left\langle \Gamma _{25^{\prime u}}\left\vert \mathbf{%
p}\right\vert \Gamma _{2^{\prime l}}\right\rangle$ &0.2475-3.0505$x$+2.6856$x^{2}$  \\
$P_{3}=\frac{\hbar }{m}\left\langle \Gamma _{25^{\prime l}}\left\vert \mathbf{%
p}\right\vert \Gamma _{2^{\prime u}}\right\rangle$ & -0.102+1.8342$x$-1.7302$x^{2}$ \\
$P_{4}=\frac{\hbar }{m}\left\langle \Gamma _{25^{\prime u}}\left\vert \mathbf{%
p}\right\vert \Gamma _{2^{\prime u}}\right\rangle$ & 1.465+0.3021$x$-0.5864$x^{2}$ \\
$Q_{1}=\frac{\hbar }{m}\left\langle \Gamma _{25^{\prime l}}\left\vert \mathbf{%
p}\right\vert \Gamma _{15}\right\rangle$ & 1.1275-0.1851$x$-0.0077$x^{2}$ \\
$Q_{2}=\frac{\hbar }{m}\left\langle \Gamma _{25^{\prime u}}\left\vert \mathbf{%
p}\right\vert \Gamma _{15}\right\rangle$ & -0.7412+4.7825$x$+3.4247$x^{2}$   \\
$R_{1}=\frac{\hbar }{m}\left\langle \Gamma _{25^{\prime l}}\left\vert \mathbf{%
p}\right\vert \Gamma _{12^{\prime }}\right\rangle$ & 0.5220+0.0472$x$+0.0124$x^{2}$  \\
$R_{2}=\frac{\hbar }{m}\left\langle \Gamma _{25^{\prime u}}\left\vert \mathbf{%
p}\right\vert \Gamma _{12^{\prime }}\right\rangle$ & 0.9477-1.6317$x$ \\
$T_{1}=\frac{\hbar }{m}\left\langle \Gamma _{1^{u}}\left\vert \mathbf{p}%
\right\vert \Gamma _{15}\right\rangle $ &1.1108-0.0786$x$-0.0613$x^{2}$  \\
$T_{2}=\frac{\hbar }{m}\left\langle \Gamma _{1^{l}}\left\vert \mathbf{p}%
\right\vert \Gamma _{15}\right\rangle $ & -0.0534+0.5107$x$ \\
\end{tabular}
\end{ruledtabular}
\end{table}

%\begin{table*}
%\caption{\label{tab:table3}Eigenvalues and SO splitting of $\Gamma$-centered states. All energies are in eV.}
%\begin{ruledtabular}
%\begin{tabular}{ccc}
%States at $\Gamma$&Ge$_{1-x}$Sn$_{x}$(MBJ) &Ge$_{1-x}$Sn$_{x}$(HSE)\\ \hline
%$\Gamma_{1^{l}}$ &-12.2519+1.4249$x$ &-12.2519+0.3989$x$\\
%$\Delta_{25'^{l}}$&0.2247-0.3727$x$+0.8$x^{2}$ &0.2247-0.3047$x$+0.8$x^{2}$ \\
%$\Gamma_{25'^{l}}$&0.00 &0.000  \\
%$\Gamma_{15}$ &2.990-0.796$x$ &2.990-0.468$x$ \\
%$\Delta_{15}$&0.2520+0.193$x$ &0.2520+0.23$x$ \\
%$\Gamma_{2'^{l}}$&0.8140-3.34$x$+2.15$x^{2}$ &0.8140-3.353$x$+1.96$x^{2}$\\
%$\Gamma_{1^{u}}$&8.2064-2.7334$x$ &8.2064-2.8884$x$ \\
%$\Gamma_{12'}$& 8.5786-0.9856$x$&8.5786-0.2666$x$ \\
%$\Gamma_{25'^{u}}$&13.4041-4.8581$x$ &13.4041-4.6381$x$\\
%$\Delta_{25'^{u}}$&0.0793-0.0333$x$ &0.0793-0.0323$x$\\
%$\Gamma_{2'^{u}}$&17.0426-5.5226$x$ &17.0426-5.5226$x$ \\
%\end{tabular}
%\end{ruledtabular}
%\end{table*}

\begin{table}
\caption{\label{tab:table3}Energy levels and SO splitting of $\Gamma$-centered states. All energies are in eV.}
\begin{ruledtabular}
\begin{tabular}{cc}
States at $\Gamma$&Ge$_{1-x}$Sn$_{x}$\\ \hline
$\Gamma_{1^{l}}$ &-12.2519+1.4249$x$ \\
$\Delta_{25'^{l}}$&0.2247+5.3808$x$-4.9535$x^{2}$  \\
$\Gamma_{25'^{l}}$&0.00   \\
$\Gamma_{15}$ &2.990-0.796$x$ \\
$\Delta_{15}$&0.2520+0.193$x$  \\
$\Gamma_{2'^{l}}$&0.8140-3.4667$x$+2.2767$x^{2}$ \\
$\Gamma_{1^{u}}$&8.2064-2.7334$x$ \\
$\Gamma_{12'}$& 8.5786-0.9856$x$ \\
$\Gamma_{25'^{u}}$&13.4041-4.8581$x$ \\
$\Delta_{25'^{u}}$&0.0793-0.0333$x$ \\
$\Gamma_{2'^{u}}$&17.0426-5.5226$x$ \\
\end{tabular}
\end{ruledtabular}
\end{table}

\begin{table}
\caption{\label{tab:table4}The fitting expression for band gap of $\Gamma$-valley and L-valley, effective mass of electron and hole.
         The energy unit is eV.}
\begin{ruledtabular}
\begin{tabular}{cc}
Parameter &Ge$_{1-x}$Sn$_{x}$ \\ \hline
$Eg(\Gamma)$ & 0.814-3.467$x$+2.277$x^{2}$\\
$Eg(\mathrm{L})$ & 0.670-1.74$x$+2.862$x^{2}$\\
$m_{e}^{\Gamma}$ &0.045-0.166$x$+0.043$x^{2}$ \\
$m_{l}^{L}$&1.544-1.390$x$+4.831$x^{2}$   \\
$m_{t}^{L}$&0.085-0.063$x$+0.184$x^{2}$ \\
$m_{hh}^{[001]}$ &0.194 \\
$m_{hh}^{[011]}$&0.399-0.082$x$  \\
$m_{hh}^{[111]}$&0.558-0.048$x$ \\
$m_{lh}^{[001]}$&0.058-0.258$x$+0.214$x^{2}$ \\
$m_{lh}^{[011]}$&0.050-0.204$x$+0.121$x^{2}$ \\
$m_{lh}^{[111]}$&0.048-0.194$x$+0.112$x^{2}$\\
$m_{h,\mathrm{DOS}}$&0.215-0.130$x$+0.201$x^{2}$\\
$m_{e,\mathrm{DOS}}^{\Gamma}$&0.045-0.166$x$+0.043$x^{2}$ \\
$m_{e,\mathrm{DOS}}^{L}$&0.566-0.449$x$+1.401$x^{2}$\\
%$\gamma_{1}$&11.298+27.488$x$+193.669$x^{2}$ \\
%$\gamma_{2}$&3.153+13.777$x$+96.745$x^{2}$ \\
%$\gamma_{3}$&4.751+13.670$x$+96.078$x^{2}$   \\

\end{tabular}
\end{ruledtabular}
\end{table}

The VCA \cite{rideau2006strained} is used to extend the 30-band $k$$\cdot$$p$ model to Ge$_{1-x}$Sn$_{x}$ alloy. Although the crystal structure of Ge$_{1-x}$Sn$_{x}$ alloy does not belong to group $O^{h}$ due to the broken centrosymmetry, the group $O^{h}$ model is used in order to simplify the optimization of parameters to a reasonable level \cite{rideau2006strained}.
We have used quadratic interpolation between Ge and $\alpha$-Sn to come up with the
coupling strengths (Table \ref{tab:table2}) as well as the necessary energy levels of different bands and SOC
splitting at $\Gamma$ point (Table \ref{tab:table3}) as input parameters for the 30-band $k$$\cdot$$p$ model for each alloy
composition of Ge$_{1-x}$Sn$_{x}$. Bowing effect in both conduction band minimum(CBM) and SOC in valence band
is considered. For the coupling matrix elements, we are mainly concerned with those that couple
with the conduction and valence bands. These input parameters are optimized in order for the $k$$\cdot$$p$ model to produce bandgap results that best match those in \cite{zelazna2015electronic} and to yield effective
masses close to what were obtained in \cite{lu2012electronic}.

Fig.\ref{fig4} shows the dependence of bandgap at L-valley, $E_{g}$(L) (solid blue), and at $\Gamma$-valley, $E_{g}(\Gamma)$ (solid red) on the Sn composition $x$, calculated using the 30-band model. The dashed lines are calculated results for $E_{g}(L)$ and $E_{g}(\Gamma)$ according to the empirical quadratic interpolation expression  $E_{g}(\text{Ge}_{1-x}\text{Sn}_{x})=(1-x)E_{g}(\text{Ge})+xE_{g}(\text{Sn})-bx(1-x)$, where the bowing parameter have been determined at L and $\Gamma$-valley as $b(\mathrm{L})=0.26$ eV and $b(\Gamma)=1.8$ eV, respectively \cite{zelazna2015electronic}. In comparison, our 30-band $k$$\cdot$$p$ model produces stronger bowing effect for both L and $\Gamma$-valley bandgaps as given in Table \ref{tab:table4}
where the quadratic fitting yields $b(\mathrm{L})=2.862$ eV and $b(\Gamma)=2.277$ eV, respectively. These calculated
results match well with the experimental data (scatter points in various colours in Fig. \ref{fig4}) \cite{PhysRevB.73.125207,dybala2016electromodulation,PhysRevB.77.073202,lin2012investigation}
in terms of bandgap values and rate of bandgap decrease with Sn composition. The crossover of L
and $\Gamma$-valley occurs at the Sn composition $x=7.25\%$ according to our 30-band $k$$\cdot$$p$ model, which is in
good agreement with the empirical predication of $x=7.18\%$ \cite{zelazna2015electronic} and the experimental result of
$x\approx7.8\%$ \cite{tseng2013mid}. Some of the discrepancies can be accounted for by the different bandgaps used in
the calculations for Ge as shown in Table \ref{tab:table1}.

%\begin{figure}
%\includegraphics[width=3.2in,height=2.0in]{Fig3.eps}
%\caption{The DOS of Ge. The black line denotes the total DOS and other colors lines denote the projected DOS.}
%\label{fig3}
%\end{figure}

%\begin{figure}
%\includegraphics[width=3.2in]{Fig4.eps}
%\caption{The calculated L $\Gamma$ and X valley shifts of a uniaxially strained }
%\label{fig4}
%\end{figure}

\begin{figure}
\includegraphics[width=3.0in, height=3.0in]{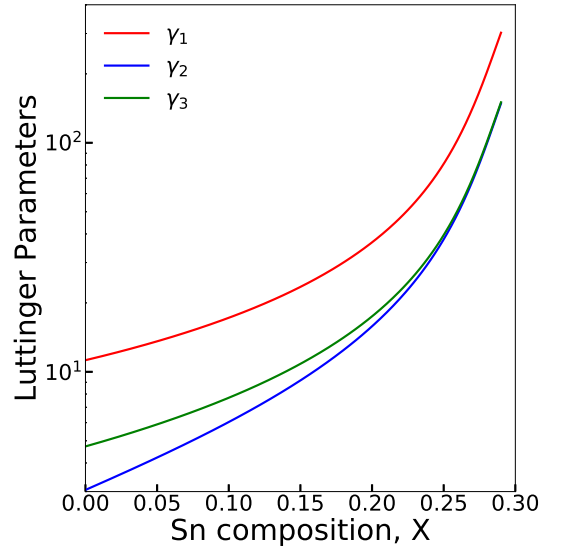}
\caption{The Luttinger parameters vs. Sn composition. The vertical axis scale is logarithmic.}
\label{fig6}
\end{figure}

\begin{figure*}
\includegraphics[width=7.2in]{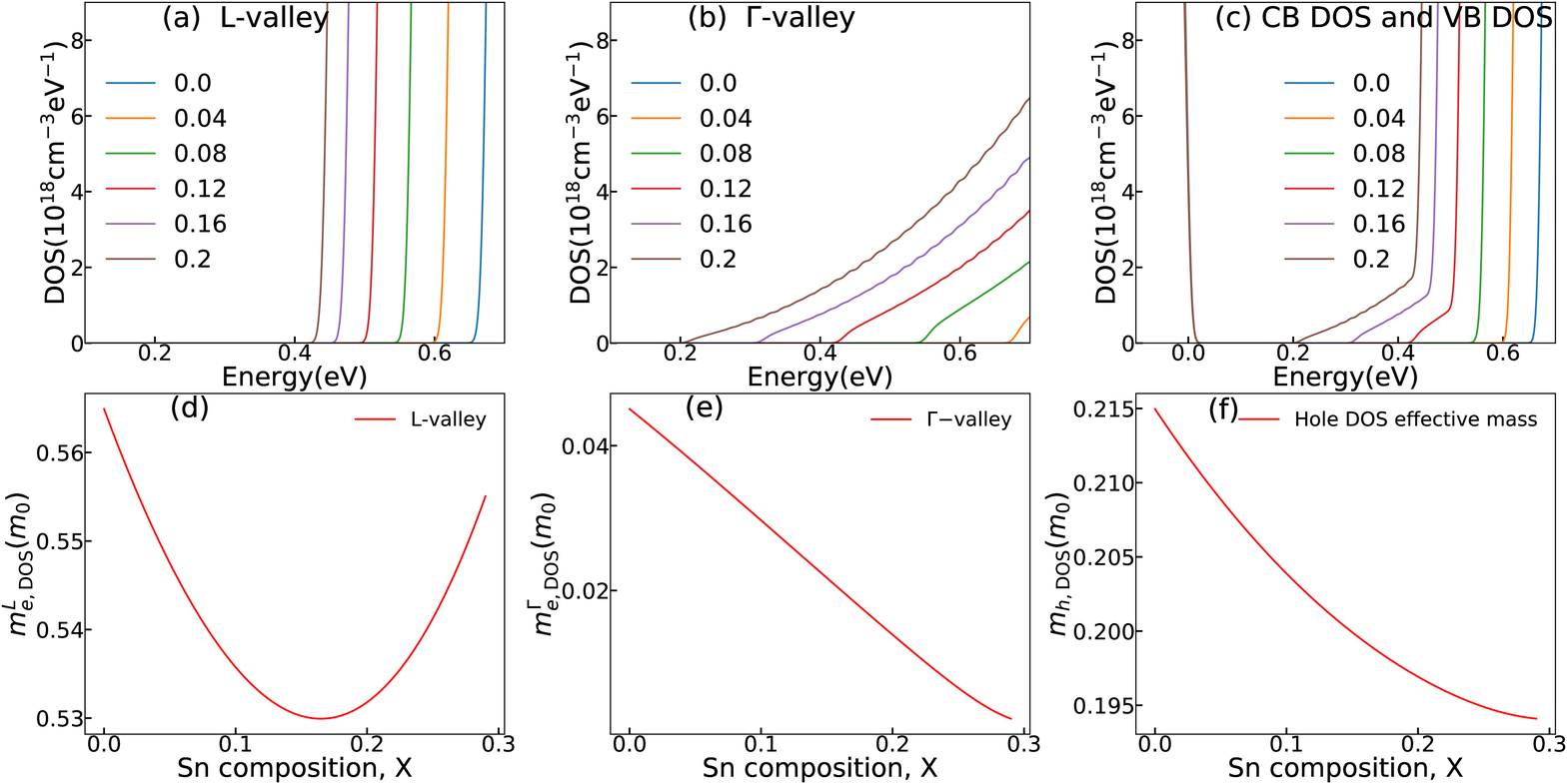}
\caption{(a)Conduction band DOS in (a) L-valley and (b) $\Gamma$-valley. (c) Combined conduction band DOS(curves in color) and valence band(solid brown). Electron DOS effective mass for (d) L-valley and (e) $\Gamma$-valley. (f) Hole DOS effective mass vs. Sn composition.}
\label{fig7}
\end{figure*}

The electronic band structure obtained from 30-band model allows for extraction of important
material parameters such as effective
mass of electrons at L and $\Gamma$-valleys, Luttinger parameters and DOS effective mass, that
directly influence electronic and optical processes in Ge$_{1-x}$Sn$_{x}$ alloy.
For such an immature material system with much to be explored, linear extrapolation of these parameters is
often used, but their accuracies are obviously unreliable, and therefore not recommended \cite{vurgaftman2003band}.
To the best of our knowledge, these important parameters have not been reported previously. Fig. \ref{fig5} shows the results of effective masses for electrons at L and $\Gamma$-valleys as well as heavy hole (HH) and light hole (LH) along different crystalline directions of the Ge$_{1-x}$Sn$_{x}$ alloy in the unit of free electron mass $m_{0}$. It can be seen from Fig. \ref{fig5}(a) and
(b)that the electron effective mass at $\Gamma$-valley is largely linear within $0<x<0.3$, while the
longitudinal and transverse electron effective masses at L-valley are highly parabolic with minima
of $m_{l}^{L}=1.44(m_{0})$ and $m_{t}^{L}=0.080({m_{0}})$ occurring at $x=14.4\%$ and $x=17.7\%$, respectively. Due to the non-parabolicity and
anisotropy of the valence band at $\Gamma$ valley, the effective masses of HH and LH
in Fig. \ref{fig5}(c) and (d) are quite different along the [001], [011] and [111] directions. Linear and
quadratic dependence of these effective masses on Sn composition $x$ are shown in Table \ref{tab:table4}.

Luttinger parameters that adequately characterize the valance band near $\Gamma$ point can also be
obtained using the following relationships with the HH and LH effective masses along [001] and
[111] directions: $\gamma _{1}=\frac{1}{2}\left( 1/m_{lh}^{[001]}+1/m_{hh}^{[001]}\right)$, $\gamma _{2}=\frac{1}{4}\left( 1/m_{lh}^{[001]}-1/m_{hh}^{[001]}\right)$, $\gamma _{3}=\frac{1}{4}\left( 1/m_{lh}^{[001]}+1/m_{hh}^{[001]}\right)-1/2m_{hh}^{[111]}$ \cite{lu2012electronic}.
As the Sn composition exceeds the crossover point, the bandgap of Ge$_{1-x}$Sn$_{x}$ alloy becomes direct,
the availability of Luttinger parameters enables the use of simpler 8-band $k$$\cdot$$p$ model as a
convenient method to investigate various phenomena that take place around the $\Gamma$ point such as
the vertical optical transitions associated with optical absorption and emission processes.
Luttinger parameters vs. the Sn composition are shown in Fig. \ref{fig6} where a strong nonlinear
dependence is revealed.

At last, DOS in conduction band and valence bands, which are key features determining a range of electronic and optical properties that impact device performance, are calculated for Ge$_{1-x}$Sn$_{x}$ alloy with different compositions.
Figure \ref{fig7} (a) and (b) show the DOS in L and $\Gamma$-valley in the conduction band as a function of the electron energy from their respective band edge, respectively. It is not difficult to see that the L-valley DOS is much higher than that in $\Gamma$-valley as expected. Figure 7(c) shows the combined conduction-band DOS which is the summation of L and $\Gamma$-valley DOS. It can be seen that as the Ge$_{1-x}$Sn$_{x}$ alloy makes the transition from indirect to direct bandgap ($x>7.25\%$), the total conduction band DOS around its band edge reduces, positively affecting all kinds of light emitting devices including LEDs and lasers in terms of their efficiency and threshold in addition to the impact of direct bandgap that facilities the optical emission process. Also shown in Fig. \ref{fig7}(c) on the left (solid brown)is the valence band DOS near its maximum in $\Gamma$-valley. The valance band DOS curves for different Sn compositions are all bunched together with no appreciable difference among them. The electron DOS effective masses at L and $\Gamma$-valley vs. Sn composition are shown in Fig. \ref{fig7}(d) and (e), respectively, where the former is consistently over an order of magnitude greater than the latter. The hole DOS effective mass shown in Fig. 7(f) varies within a narrow range of $0.195(m_{0})<m_{h,\mathrm{DOS}}<0.215(m_{0})$ which is consistent with the small variation of valence band DOS shown in Fig. \ref{fig7}(c) for different Sn compositions.

\section{Conclusion}
In conclusion, we present a 30-band $k$$\cdot$$p$ model that can be used
efficiently to calculate the electronic band structures for relaxed Ge$_{1-x}$Sn$_{x}$ alloy of various
compositions. This model was first validated to show consistency in calculating the electronic band
structures of the $\alpha$-Sn and Ge by \textit{ab initio} and ETB methods, respectively. Two sets of input
parameters were optimized for band structure calculation at various Sn compositions: 1) coupling
matrix elements and 2) eigen energy levels and SO splitting at $\Gamma$ point for various Sn compositions,
$0<x<0.3$. Based on this model, the indirect-to-direct crossover is determined to take place at the Sn
composition of $7.25\%$ which is good agreement with previous predication and experimental result.
Bowing parameters that describe the relationship between bandgaps at $\Gamma$ and L-valley and Sn
composition are obtained. The calculated band structure allows for extraction of electron effective
mass at $\Gamma$ valley, longitudinal and transverse electron effective mass at L-valley, as well as HH and
LH effective masses along different crystalline directions. From these HH and LH effective masses,
Luttinger parameters that characterize the valence band around $\Gamma$-valley in Ge$_{1-x}$Sn$_{x}$ alloy can be
established. Finally, DOS in both conduction and valence band as well as their associated effective-
mass DOS can be calculated as well. The 30-band $k$$\cdot$$p$ model fulfills the need for an efficient and
effective tool that can be used in calculating the electronic band structure of Ge$_{1-x}$Sn$_{x}$ alloy across its
entire BZ, establishing material parameters, and optimizing device performance.

\section{acknowledgements}
 WJ Fan acknowledges the funding support (NRF--CRP19--2017--01). The computation of this work was partially performed on resources
of the National Supercomputing Centre, Singapore. G Sun
acknowledges the grant support (FA9550-17-1-0354) from the Air Force Office of Scientific
Research.

\bibliography{main}

\end{document}